\documentclass[11pt,twocolumn]{aastex62}

\shorttitle{Blanco\,1 Star Cluster}
\shortauthors{Zhang et al.}

\begin{document}

\hfill Today is \today
\received{August 19, 2019}
\revised{December 6, 2019}
\accepted{December 12, 2019}
\submitjournal{ApJ}

%\title{ Characterization of the Stellar and Substellar Members 
%        in the Nearby Blanco\,1 Star Cluster }
\title{ Diagnosing the Stellar Population and Tidal Structure of the Blanco\,1 Star Cluster}

\author{Yu Zhang}
    \affiliation{Xinjiang Astronomical Observatory, Chinese Academy of Sciences, P. R. China}
    \email{zhy@xao.ac.cn}

\author{Shih-Yun Tang}
    \affiliation{Department of Physics, National Central University, 
                300 Zhongda Road, Zhongli, Taoyuan 32001, Taiwan}
 %\email{sytang@g.ncu.edu.tw}

\author{W.~P. Chen}
    \affiliation{Graduate Institute of Astronomy, National Central University, 
                300 Zhongda Road, Zhongli, Taoyuan 32001, Taiwan} 
    \affiliation{Department of Physics, National Central University, 
                300 Zhongda Road, Zhongli, Taoyuan 32001, Taiwan}
%    \email{wchen@astro.ncu.edu.tw}

\author{Xiaoying Pang}
    \affiliation{Xi'an Jiaotong-Liverpool University, 111 Ren'ai Road, Dushu Lake Science 
                and Education Innovation District, Suzhou 215123, Jiangsu Province, P. R. China}
    \affiliation{Shanghai Key Laboratory for Astrophysics, Shanghai Normal University, 
                100 Guilin Road, Shanghai 200234, P. R. China}
%    \email{Xiaoying.Pang@xjtlu.edu.cn}

\author{J.~Z. Liu}
    \affiliation{Xinjiang Astronomical Observatory, Chinese Academy of Sciences,  P. R. China}
    % \email{liujinzh@xao.ac.cn }

\correspondingauthor{Shih-Yun Tang}
\email{sytang@g.ncu.edu.tw}

%---------------------------------------------------------------------------------------%
%---------------------------------------------------------------------------------------%
\begin{abstract}
We present the stellar population, using {\it Gaia}\,DR2 parallax, kinematics, and photometry, 
of the young ($\sim 100$~Myr), nearby ($\sim 230$~pc) open cluster, Blanco\,1.  A total of 
644 member candidates are identified via the unsupervised machine learning method \textsc{StarGO} 
to find the clustering in the 5-dimensional position and proper motion parameter 
($X$, $Y$, $Z$, $\mu_\alpha \cos\delta$, $\mu_\delta$) space. Within the tidal radius of 
$10.0 \pm 0.3$~pc, there are 488 member candidates, 3 times more than those outside. 
A leading tail 
and a trailing tail, each of 50--60~pc in the Galactic plane, are found for the first time for this 
cluster, with stars further from the cluster center streaming away faster, manifest stellar stripping.
Blanco\,1 has a total detected mass of $285\pm32$~M$_\sun$ with a mass function 
consistent with a slope of $\alpha=1.35\pm0.2$ in the sense of $dN/dm \propto m^{-\alpha}$, in the mass 
range of 0.25--2.51~M$_\sun$, where $N$ is the number of members and $m$ is stellar mass. 
 A Minimum Spanning Tree ($\Lambda_{\rm MSR}$) analysis shows the cluster to be moderately mass 
segregated among the most massive members ($\ga 1.4$~M$_\sun$), suggesting an early stage of dynamical 
disintegration.
\end{abstract}
%---------------------------------------------------------------------------------------%
%---------------------------------------------------------------------------------------%

%sec1
\section{Introduction} \label{sec:intro}

Star formation takes place in dense molecular clouds. While individual 
stars are formed in dense cores, collectively a giant molecular cloud produces a complex of 
star clusters \citep{lad03}. 
The shape of a star cluster bears the imprint of its formation and evolutionary history. 
At birth the stellar distribution inherits the generally filamentary structure of the 
parental molecular cloud \citep{che04}. Thereafter, through mutual gravitational interaction between 
member stars, higher-mass stars lose kinetic energy and sink to the center, whereas 
lower-mass members gain speed, thereby occupying a progressively larger volume of space. Those 
low-mass members at the outermost region are vulnerable to external forces, e.g., the  
differential rotation, disk shocks, spiral arm passage, etc., leading to tidal structures 
containing escaping members. 

Even halo globular clusters, while spending much of their lifetime in relative isolation in the 
Galactic halo, are also elongated, averaging an aspect 
ratio of 0.87, which cannot be accounted for by rotation, but could be attributed 
mainly to the tidal stretching by the bulge, manifest by 
the protrusion of globular clusters in the vicinity of the Galactic center, some 
with possible stellar debris \citep{che10}.  The most notable example of a disrupting globular 
cluster is perhaps Palomar~5, which is known to have tails spanning symmetrically on either
side of the cluster \citep{ode01}, with the latest studies revealing 
an extent more than 20~deg \citep{kuz15}. The tails contain more stars than the cluster itself 
\citep{ode01,ode03}, implying an advanced stage of cluster disintegration. The $N$-body 
simulations conducted by \citet{deh04} lend support to disk crossing being the primary 
mechanism for creation of the tail structure, and predicted a likely destruction of the 
cluster in its next disk crossing event in about 110~Myr. 

%A fluid experiencing a gravitational tidal force bulges up, resulting a tail and antitail 
%pointing toward the gravitating body. The unbound members are vulnerable to be stripped away, 
%leaving behind debris stars in the trailing tail along the cluster orbit. For a compact gravitating
%source, as in the case for the bulge on globular clusters, the tidal effects include stretching in 
%the direction toward the source and slight compression perpendicular 
%to this direction.  For open clusters near the plane, where the local disk surface density 
%varies smoothly, there is little perpendicular compression.}

It has been challenging to recognize such tidal tails for open clusters because of the difficulty 
in distinguishing members in the tails from field stars.
However, with the {\it Gaia} data release~2 (DR2) availing high-precision photometry, proper 
motion (PM) and parallax data, detection of tidal tails in open 
clusters in the solar neighborhood has been mushrooming
(Hyades: \citet{ros19a,mei18}, Coma Berenices: \citet{fur19,tan19}, Praesepe: \citet{ros19b}). 
These clusters are all relatively old, $\sim600$--800~Myr \citep{ros19a,tan19,ros19b}, and
located away from the Galactic Plane 
(Hyades: $\ell=179\fdg9184$, $b=-20\fdg6883$,
Coma Berenices: $\ell=220\fdg9594$, $b=+83\fdg7630$, and
Praesepe: $\ell=205\fdg8970$, $b=+32\fdg4712$), 
which makes face-on structures readily detected and characterized.

Blanco\,1 ($\ell=15\fdg5719$, $b=-79\fdg2612$) has a relatively young age of $\sim 100$~Myr 
\citep{pla11}, a heliocentric distance of $\sim 237$~pc \citep{gai18b}, and is located toward 
the South Galactic Pole.  \citet{bla49} discovered the cluster by noticing an over density 
of A0 type stars in the vicinity of $\zeta$ Sculptoris.  The cluster has been investigated 
in photometry \citep{per78, ded85, wes88}, radial velocity \citep{mer08,gon09}, X-ray emission  
\citep{mic99, pil03, pil04}, and PM \cite{pla11}.  With a few hundreds of possible members 
\citep{gai18b} distributed in a projected angular size of 4$\degr$, Blanco\,1 is relatively 
sparse and thus, despite its proximity, has not been well studied as other nearby star clusters. 
No tidal structure has ever been reported for this star cluster.

In this work, using {\it Gaia}/DR2 data, we present detailed characterization of the Blanco\,1 
star cluster, by identification of its members, with which the cluster parameters, including 
tidal structures, are derived. In Section~\ref{sec:data}, we describe the {\it Gaia}/DR2 
data used, the quality control procedure, and the methodology of the member selection.
Section~\ref{sec:discussion} reports on the age, morphology, cluster mass, and mass segregation 
of the cluster based on the member list. A summary is outlined in Section~\ref{sec:summary}.

%---------------------------------------------------------------------------------------%
%---------------------------------------------------------------------------------------%
\section{Data and Analysis}\label{sec:data}
%-------------------------------------------------------------------------------------------%
\subsection{Gaia\,DR2 Data Processing and Analysis}\label{sec:gaia}

The DR2 of the ongoing {\it Gaia} space mission provides a catalog of approximately 1.3 billion
sources with high-quality photometry, PMs ($\mu_\alpha \cos\delta, \mu_\delta$), and parallaxes 
($\varpi$) \citep{gai18}. Sources with $G$ magnitudes $\leq14$, $17$ and $20$~mag 
have typical PM uncertainties of $0.05$, $0.2$ and $1.2$~mas~yr$^{-1}$, and 
$\varpi$ uncertainties of $0.04, 0.1$, and $0.7$~mas, respectively.  Typical photometric 
uncertainties at $G=$17~mag are $\Delta G=2$~mmag, $\Delta G_{\rm BP}=10$~mmag, 
and $\Delta G_{\rm RP}=10$~mmag. In this study, to exclude possible
artifacts, we apply the quality cut suggested by \citet{lin18} (see Appendix~\ref{sec:qulitycut}).

\begin{figure}[tb!]
\centering
\includegraphics[angle=0, width=1.\columnwidth]{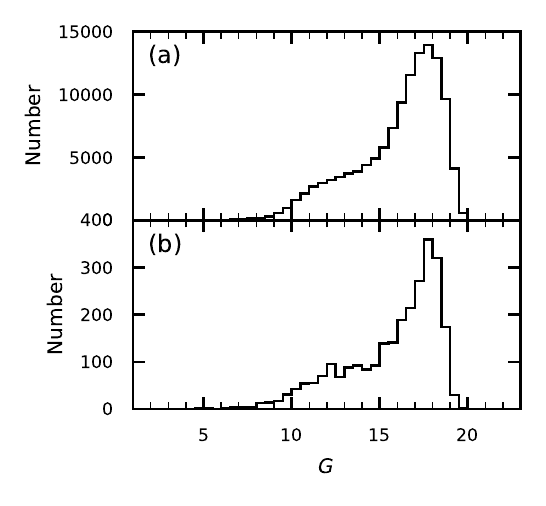}
\caption{ 	 
        Number histogram of {\it Gaia} DR\,2 stars in $G$ magnitudes for (a) Sample~I and for 
        (b) Sample~II (see text.)
	    }
\label{fig:ghist}
\end{figure}

\begin{figure*}[htb!]
\centering
\includegraphics[angle=0, width=1.\textwidth]{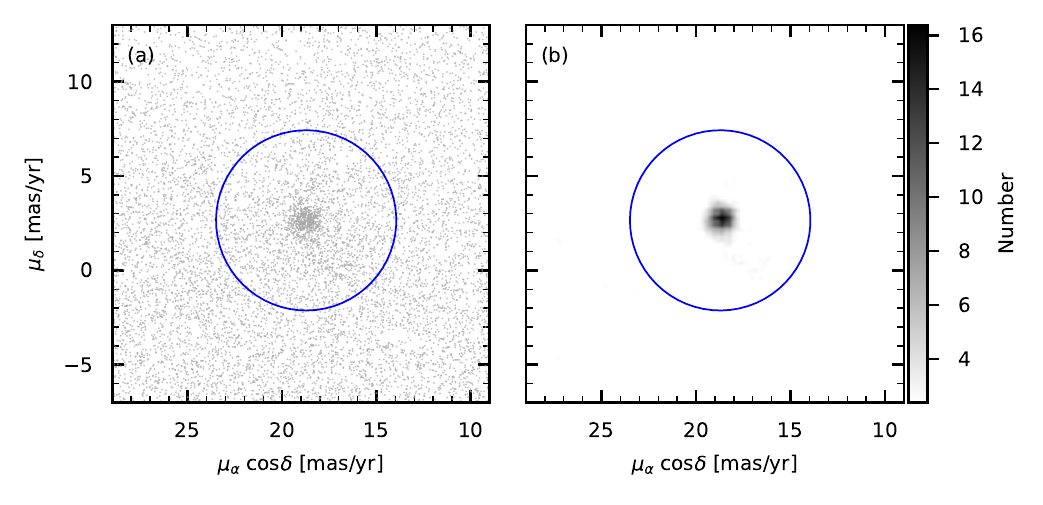}
\caption{ 	
        (a) Proper motion vector plot for all stars toward Blanco\,1 (Sample~I). 
	    (b) 2D density map of (a). Each bin is smoothed by 
	        neighboring 8 bins and only bins with a number count above  
	        2.4 (3$\sigma$, where $\sigma$ is the standard deviation of all bins) are shown.
	    The blue circle, with a radius of 6$\sigma$ ($=4.8$~mas~yr$^{-1}$), marks the PM 
	    selection range for Sample~II. 
	    }
\label{fig:pmhist}
\end{figure*}

Data were processed similar to the procedure described in \citet{tan19}.  First, a radius 
of 100~pc centering around the Galactocentric coordinates ($X,Y,Z$) = ($-8256.7, +11.4, -205.9$)~pc 
of Blanco\,1 is selected, with the coordinates transformed from the R.A., Decl., and parallax adopted 
from \citet{gai18b} via the Python \texttt{Astropy} package \citep{ast13,ast18} \footnote{Assumptions 
on the Galactocentric coordinates transformation are summarized in Appendix~A of \citet{tan19}}.
This sample contains 124,137 sources, and is called Sample~I, with the $G$ magnitudes ranging from 
$\sim4.5$ to $\sim 20.4$~mag, and with a distribution function turning down, i.e., being significantly 
incomplete, beyond $\sim18.5$~mag, shown in Figure~\ref{fig:ghist}~(a). A further selection was done on
the basis of the PM. Figure~\ref{fig:pmhist}~(a) displays the PMs of Sample~I, with the 2-dimensional 
histogram presented in Figure~\ref{fig:pmhist}~(b).  A concentration is clearly seen with $>3\sigma$ 
significance. Stars within a radius of 4.8~mas~yr$^{-1}$ (i.e., 6$\sigma$) from the PM center of 
($\mu_\alpha \cos\delta$, $\mu_\delta$) =  (+18.72, +2.65)~mas~yr$^{-1}$ \citep{gai18b} were 
then selected, forming Sample~II, which has 2673 stars with $G$ magnitudes ranging from $\sim 4.5$ to 
$\sim 19.8$~mag, and is also incomplete beyond $\sim 18.5$~mag, as in Sample~I, shown in 
Figure~\ref{fig:ghist}~(b). 

We incorporated the 5D parameters (R.A., Decl., $\varpi$, $\mu_\alpha \cos\delta$, and $\mu_\delta$)
from {\it Gaia}/DR2 to select member candidates. Because only a minor fraction of stars in Sample~II 
have radial velocity measurements (RVs) with sufficiently good quality (errors less than 2~km~s$^{-1}$),
the RV data therefore served only as complementary in analysis and are not used in member selection.
The distance used in this study is taken as 1/$\varpi$, since all stars in Sample~II are
within 350~pc from the Sun, leading to an expected distance difference between 1/$\varpi$ and that 
in the \citet{bai18}'s catalog by only about 3~pc, the correction of the global parallax
zeropoint of $\sim0.03$~mas \citep{lin18}. Using 1/$\varpi$ as distance, we computed  
for each source the Galactocentric Cartesian coordinates ($X, Y, Z$). 

%fig1

%fig2

%---------------------------------------------------------------------------------------%

\subsection{Member Selection}\label{sec:selection}

\begin{figure*}[tb!]
\centering 
\includegraphics[angle=0, width=1.\textwidth]{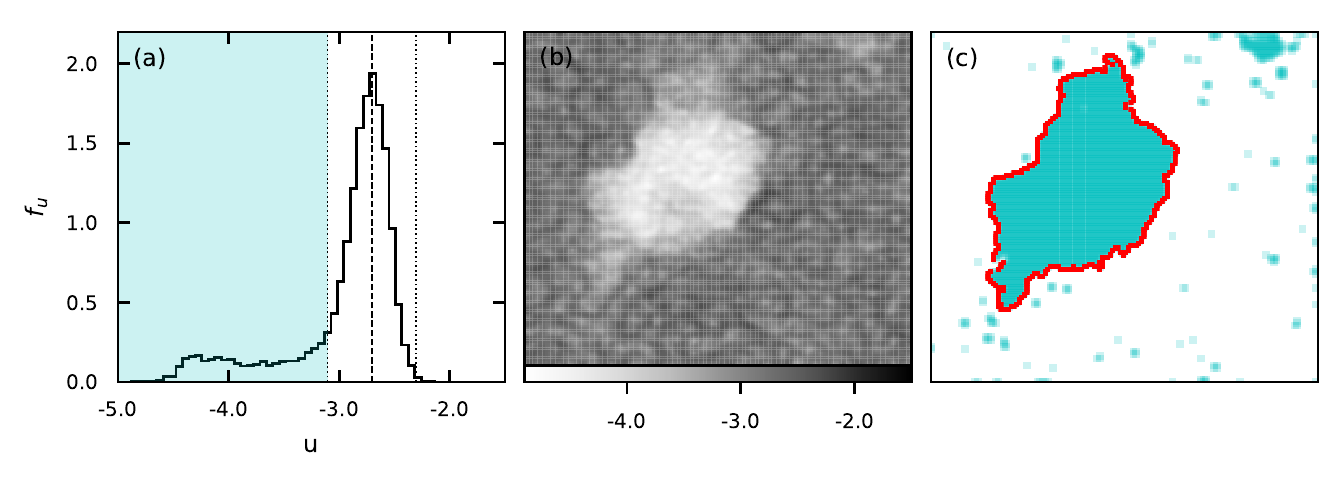}
\caption{Stellar group identified by \textsc{StarGO} in the 5D-parameter
        ($X, Y, Z$, $\mu_\alpha \cos\delta, \mu_\delta$) space. 
	    (a) Distribution histogram of $u$. The dashed line and the dotted lines denote, 
            respectively, the peak position ($u_{\mathrm{peak}}$), and the the 3$\sigma$ range, 
           ($u_{\mathrm{peak}\pm 3\sigma}$).
	        The part of $u<u_{\mathrm{peak}-3\sigma}$ is highlighted in cyan.
    	(b) 2D neural map resulting from SOM, where the $u$ value between adjacent neurons 
    	    is represented by grayscale. 
	    (c) The same as in (b) but with the neurons with $u<u_{\mathrm{peak}-3\sigma}$ colored
	        in cyan.  The red contour  traces the dominant neuron group.
	   }
\label{fig:som}
\end{figure*}

We applied an unsupervised machine learning method,
\textsc{StarGO}\footnote{\url{https://github.com/salamander14/StarGO}} \citep{yua18} 
to select member candidates. This method is built with the Self-Organizing-Map to map a 5D 
data set ($X, Y, Z$, $\mu_\alpha \cos\delta, \mu_\delta$) onto a 2D neural network, with the
topological structures of the data being preserved during dimension reduction. Thus, stars
clustered in the 5D space are associated with the neurons grouped in the 2D map. A detail 
description of \textsc{StarGO} can be found in \citet{yua18}, and an application of the member
section of the Coma Berenices star cluster is in \citet[their Section~2.3]{tan19}. In brief, we started 
out with a $150\times150$ network, with each neuron having a weight vector with the same dimension
as the input vector. We then ingested stars from our sample one by one to all the 22,500 neurons.
Each neuron would update the weight vector to become closer to the input vector of a particular
star. One iteration was complete after the neurons were trained by all stars in Sample~II once, 
and the whole learning process was iterated 400 times when the weight vectors reached convergence. 
We visualize the trained neural network by Figure~\ref{fig:som}~(b) showing the difference of
weight vectors between adjacent neurons, which is denoted by $u$.  Note that the lesser $u$ is,
the more similar the 5D parameters of the adjacent neurons are. 

Patches with lighter shades in Figure~\ref{fig:som}~(b) signify over-densities in the input 5D 
parameters. One such patch was further identified by selecting the extended distribution of $u$ in 
Figure~\ref{fig:som}~(a). We first located the peak position $u_{\mathrm{peak}}$ and the 99.85 
percentile of the distribution $u_{99.85\%}$, denoted by the dashed line and dotted lines, 
respectively, in Figure~\ref{fig:som}~(a)). The difference, $u_{99.85\%}-u_{\mathrm{peak}}$, is 
equivalent to the 3$\sigma$ confidence interval of a normal distribution, which is denoted as 
$\Delta_{3\sigma}$. The distribution $u_{\mathrm{peak-3\sigma}}$ = $u_{\mathrm{peak}}-\Delta_{3\sigma}$ 
is shown as the cyan area in Figure~\ref{fig:som}~(a), and the corresponding neurons are represented
by cyan pixels in Figure~\ref{fig:som}~(c), with the grouping of stars of Blanco\,1 enclosed by 
a red contour. 
 The over-density patch seen to the upper right corner contributes to the faint extension 
out to $\mu_\delta\sim -2.5$~mas~yr$^{-1}$ in Figure~\ref{fig:pmhist} but is 
found not to be spatially connected with the cluster. 
At the end, a total of 644 stars are selected as member candidates.
Table ~\ref{tab:blanco1} lists these candidates with the first column being the running number, 
followed by the Gaia/DR2 data (position, $\varpi$, PM, RV, $G$ magnitude and respective associated
errors) from columns 2 to 12, and a remark in column 13 of whether a star is considered within the 
tidal radius of the cluster or beyond.  

% fig3

\startlongtable
\begin{deluxetable*}{r cC c cc c Cc c Cc c Cc c cc}
\tablecaption{Blanco~1 Member Candidates\label{tab:blanco1}
             }
\tabletypesize{\scriptsize}

\tablehead{ 
	 \colhead{No.}		                     & \colhead{R.A.}                & \colhead{Decl.}                   &&
	 \colhead{$\varpi$}                      & \colhead{$\Delta \varpi$} 	     &&
	 \colhead{$\mu_\alpha \cos\delta$}       & \colhead{$\Delta(\mu_\alpha \cos\delta)$} &&
	 \colhead{$\mu\delta$}                   & \colhead{$\Delta\mu\delta$}             &&
     \colhead{RV}                            & \colhead{$\Delta$RV}               &&
     \colhead{$G$}                           & \colhead{Remark$^\dagger$}                                         \\
	 \colhead{}                              & \multicolumn{2}{c}{(J2015.5 deg)}                                 &&
	 \multicolumn{2}{c}{(mas)}               &&
	 \multicolumn{2}{c}{(mas~yr$^{-1}$)}     &&
	 \multicolumn{2}{c}{(mas~yr$^{-1}$)}     &&
	 \multicolumn{2}{c}{(km~s$^{-1}$)}       &&
     \colhead{(mag)}                         &  \colhead{}                                                       \\
     \cline{2-3} \cline{5-6} \cline{8-9} \cline{11-12} \cline{14-15}
     \colhead{(1)} & \colhead{(2)} & \colhead{(3)} && \colhead{(4)} & \colhead{(5)} && \colhead{(6)} & \colhead{(7)} &&
     \colhead{(8)} & \colhead{(9)} && \colhead{(10)} & \colhead{(11)} && \colhead{(12)} & \colhead{(13)}
	 }
%\colnumbers
\startdata
1   & 0.017250 & -29.001729 && 4.66 & 0.30 && 19.74 & 0.33 && 2.15 & 0.30 && \nodata & \nodata && 18.31 & b \\
2   & 0.028883 & -29.749518 && 4.04 & 0.14 && 18.26 & 0.27 && 2.48 & 0.20 && \nodata & \nodata && 17.45 & b \\
3   & 0.038357 & -30.091787 && 4.01 & 0.25 && 18.94 & 0.23 && 2.94 & 0.24 && \nodata & \nodata && 17.99 & b \\
4   & 0.073922 & -30.743971 && 4.02 & 0.14 && 18.23 & 0.24 && 2.86 & 0.18 && \nodata & \nodata && 17.84 & b \\
5   & 0.083604 & -29.939293 && 4.54 & 0.17 && 17.42 & 0.25 && 3.02 & 0.18 && \nodata & \nodata && 16.83 & b \\
489 & 0.194265 & -29.134891 && 3.40 & 0.05 && 19.75 & 0.08 && 2.88 & 0.06 && 7.12    & 0.56    && 10.89 & t \\
490 & 0.555670 & -27.067179 && 4.29 & 0.07 && 19.22 & 0.08 && 2.65 & 0.06 && 6.58    & 0.33    && 10.13 & t \\
491 & 1.309033 & -34.949547 && 4.09 & 0.05 && 19.26 & 0.10 && 4.03 & 0.06 && 7.75    & 2.14    && 10.82 & t \\
492 & 1.310961 & -34.948232 && 4.06 & 0.11 && 19.07 & 0.22 && 3.53 & 0.15 && \nodata & \nodata && 16.85 & t \\
493 & 1.446814 & -32.589581 && 4.35 & 0.17 && 18.84 & 0.27 && 2.89 & 0.24 && \nodata & \nodata && 18.03 & t \\
\enddata
\tablecomments{
    Entries are sorted according to R.A. in column 2. 
    This table is available in its entirety in a machine-readable form in the online journal.
    Here we only show the first five member candidates in the bound (within tidal radius) and in 
    the tail regions.\\
    $^\dagger$b: A ``bound'' member candidate within the tidal radius; t: A member candidate in the ``tail''
    (see Section~\ref{sec:tail})
    }
\end{deluxetable*} 

%---------------------------------------------------------------------------------------%

Our results are directly relevant to the work by \citet{gai18b}, using also the {\it Gaia}/DR2
data. Of the 489 member candidates they found, 427 are also in our list of candidates. In
general, we exercised a slightly different set of membership criteria, e.g., on the photometric 
signal-to-noise ratios than in \citet{gai18b}. On the other hand, we imposed a volume limit of
a 100-pc radius around the cluster center, and allowed \textsc{StarGO} to find grouping.  
This is much larger than the search radius used by \citep{gai18b}, and enabled us to recognize 
the tail structure. Indeed a significant fraction of our candidates are located in the tails, 
by missed by \citet{gai18b}.  

There are 62 candidates found by \citet{gai18b} but not in our list. The majority of these did not
pass our selection because of their inferior photometric quality. It is worth noting that except two, 
none of these stars pass the selection rules listed in \citet[][ their Appendix B]{gai18b}. 
At the moment, we could not resolve the controversy.
   
\begin{figure*}[tb!]
\centering 
\includegraphics[angle=0, width=1.\textwidth]{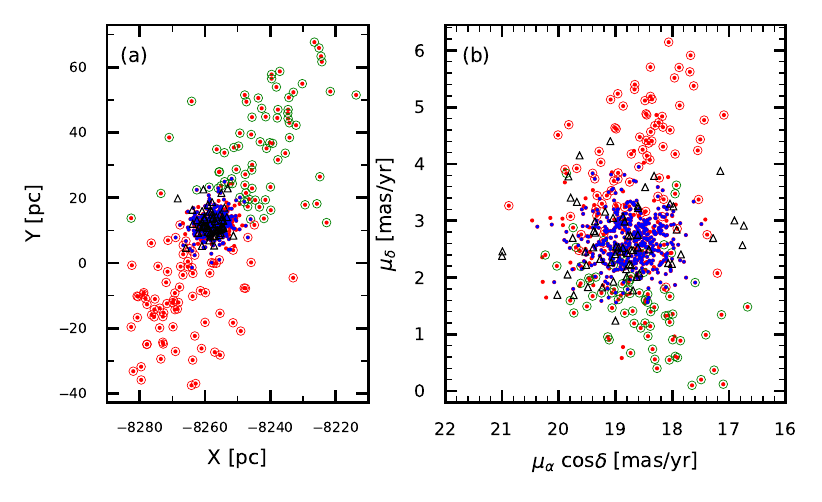}
\caption{
        (a) The Galactocentric Cartesian coordinates and 
        (b) the proper motion distributions of member candidates of Blanco\,1. 
        The colored dots represent the 644 \textsc{StarGO} member candidates, 
        of which the blue dots mark the 427 candidates also found by \citet{gai18b}, 
        whereas the red dots represent those found only by \textsc{StarGO}.  
        The 62 member candidates found only by \citet{gai18b} are marked with black 
        open circles.  The green and red open circles superimposed 
        on the dots represent member candidates on each side of the cluster.  
        }
\label{fig:gaiavsstargo}
\end{figure*}

Figure~\ref{fig:gaiavsstargo} plots the spatial and PM distributions for the member candidates 
of Blanco\,1 reported here and those by \citet{gai18b}.  Our candidates span a wider space and 
proper motion ranges, covering the tails. The 62 ``missing'' candidates reported by 
\citet{gai18b} are mostly faint (Figure~\ref{fig:cmd}~(a)).  They may still be possible candidates, but we do not include 
them in our list.  

%While the \citet{gai18b} sample is more concentrated, we note that many of 
%the 62 ``missing'' candidates are outliers, especially in the PM distribution. 

%fig4

\begin{figure}[tb!]
\centering 
\includegraphics[angle=0, width=0.9\columnwidth]{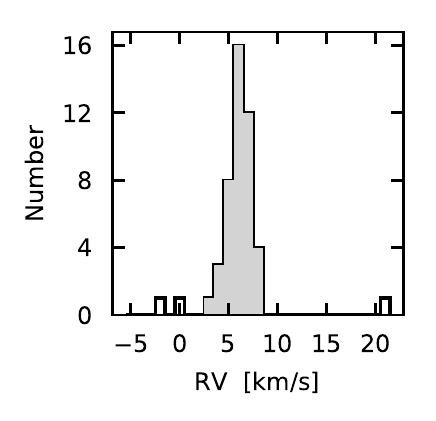}
\caption{ 
        The {\it Gaia}/DR2 radial velocity distribution of 47 member 
        candidates. The concentration (in gray), after the 3 outliers 
        are excluded, is consistent with RV=$6.1\pm1.1$~km~s~$^{-1}$ with 
        the error being the standard deviation of individual measurements.  
        }
\label{fig:rv}
\end{figure}

Of our 644 member candidates, {\it Gaia}/DR2 provides 82 RV measurements, among which 47 have
errors less than 2~km~s$^{-1}$. Even though RV is not used in our membership selection,
Figure~\ref{fig:rv} shows a clear concentration in the distribution, with 
RV=$6.1\pm1.1$~km~s$^{-1}$. This sub-sample excludes possible binary systems, and represents 
the average RV of the star cluster.  

\subsection{Contamination Rate}\label{sec:contam}

With parallax information, there should be essentially little foreground or background
stellar contamination.  The only possible contaminants remain the field stars inside
the cluster volume with similar PMs to those of cluster members. We estimated this 
contamination from the smooth Galactic disk population with the {\it Gaia}/DR2 mock catalog 
\citep{ryb18}, by applying the same spatial and PM criteria as described in Section~\ref{sec:gaia}.
This led to a mock Sample~II of 2090 stars. Analysis of this mock Sample~II with the same procedure
resulted in 24 stars associated with Blanco\,1 which, given the size of the mock Sample~II (2090) 
relative to Sample~II (2673), led to about 31 possibly field stars inside the cluster region. 

%Sample~II (2673 stars) is 1.28 times of the mock Sample~II, 
%the contamination fraction is estimated to be about $\sim1$\% ($24\times1.28/2673$)??.

%---------------------------------------------------------------------------------------%
%---------------------------------------------------------------------------------------%
\section{Discussion}\label{sec:discussion}
%---------------------------------------------------------------------------------------%
%---------------------------------------------------------------------------------------%

\subsection{The Cluster Age}\label{sec:cmd}

With ample X-ray emission \citep{pil04,pil05}, H$\alpha$ emission \citep{pan97}, and 
lithium-bearing members \citep{sta98}, Blanco\,1 has been known to be a young system. Lacking
members beyond the main sequence turn-off, however, the cluster has an uncertain age 
determination, ranging from 80~Myr using main sequence fitting \citep{car09}, $90\pm25$~Myr
using the fraction of H$\alpha$ emission-line stars \citep{pan97}, 125~Myr using lithium 
depletion analysis \citep{sta98}, to $146^{+13}_{-14}$~Myr using gyrochronology.  

\citet{gai18b} derived a logarithmic age (years) of 8.30 ($\approx200$~Myr), using photometry
of member candidates to fit the main sequence with PARSEC isochrones assuming a metallicity of 
${\rm [Fe/H]}$=0.04 \citep{for05} (i.e., $Z=0.017$). At the end, a logarithmic age of 8.06 
(=115~Myr) was adopted, on the basis of the work by \citet{jua14} who estimated the age by 
the lithium-depletion boundary, i.e., the transition among member stars from showing lithium in
the spectra to being fully depleted. This is consistent with the age of 125~Myr previously
derived also with the lithium-depletion boundary by \citet{sta98}.    

%We checked the 14 members listed in Table~1 of \citet{jua14} that consists of bright 
%stars selected from the B1opt-SMARTS optical survey, and faint ones  
%from the CFHT-BL optical survey by \citet{mor07}.  

We checked the 14 members listed in Table~1 of \citet{jua14}, which consists of 4 
bright stars selected from the B1opt-SMARTS optical survey and 10 faint stars from the 
CFHT-BL optical survey \citep{mor07}. For the bright sample, we could not match any 
counterpart of B1opt-6335, but otherwise the rest three, 
B1opt-18229 (2MASS\,J00013984$-$3004383), 
B1opt-2156 (2MASS\,J00074089$-$3005571), and 
B1opt-13328 (2MASS\,J00042277$-$3023064) have parallax and proper motion measurements consistent 
with membership of Blanco\,1, and indeed they are included in our member list.  

All the 10 stars in the faint sample ($G \leq 20$~mag) of \citet{jua14} have spectral 
types later than M5, so were considered low-mass stars or brown dwarfs. We note that 
two stars had their coordinates erroneously passed on by \citet{jua14} from \citet{mor07}; 
those for CFHT-BL$-$25 should have been 
R.A.=00:00:42.754, Decl.=$-$30:17:43.74 (J2000), and those for CFHT-BL$-$36 should have been 
R.A.=00:00:08.811, Decl.=$-$30:06:42.53 (J2000). 

Furthermore, all these 10 stars have either no {\it Gaia} measurements 
(CFHT-BL$-$22, CFHT-BL$-$29, CFHT-BL$-$45, CFHT-BL$-$49), or are uncertain in membership 
because of their relatively large errors in parallax or proper motion measurements.
%(CFHT-BL$-$16, CFHT-BL$-$24, CFHT-BL$-$25?, CFHT-BL$-$38, CFHT-BL$-$43, 
%CFHT-BL$-$36).  While the {\it Gaia} data may not be reliable because 
%of their faintness, this leaves only one star, CFHT-BL$-$25
In any case, because age determination by lithium depletion or by H-alpha emission relies on
a reliable and complete list of cluster members, the age analysis of Blanco\,1 by these methods
should be revisited.

\begin{figure*}[tb!]
\centering 
\includegraphics[angle=0, width=1.\textwidth]{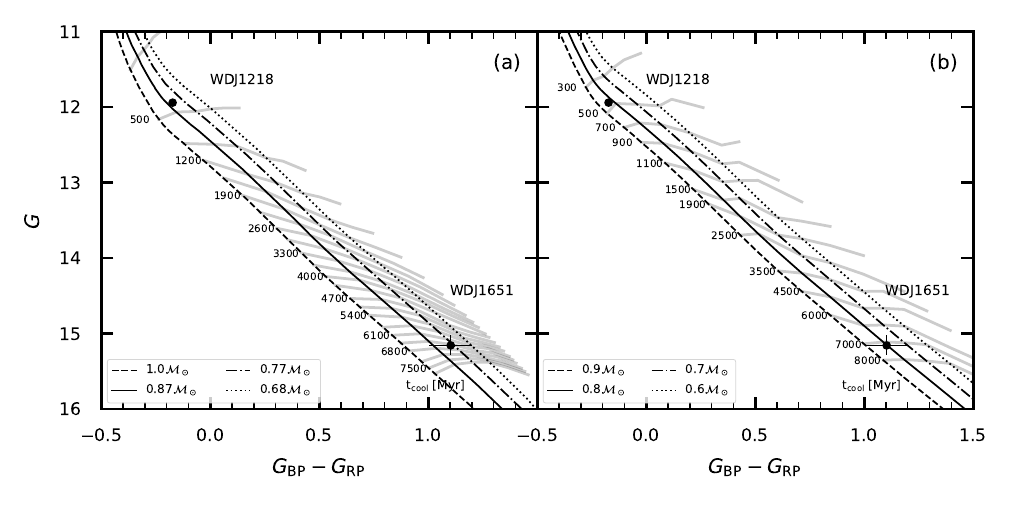}
\caption{The {\it Gaia}/DR2 CMD for the two white dwarfs, represented as dots, 
   in the  Coma Berenices field.  Also shown are the cooling curves (a)~from \citet{sal10} 
   and (b)~from \citet{tre11} at different ages (solid gray curves in Myr) 
   for different masses.  
        }
\label{fig:wd_coma}
\end{figure*}

The age of a star cluster can be constrained also by the cooling timescales of member white
dwarfs. To validate the technique, we first applied it to the  Coma Berenices star 
cluster \citep[age 700--800~Myr]{tan19}. Figure~\ref{fig:wd_coma} compares the color-magnitude
behavior of the two white dwarfs seen toward the cluster with theoretical cooling models
\citep{sal10,hol06,tre11,ber11} \footnote{For tabulation, see 
\url{http://basti.oa-teramo.inaf.it/index.html}, and 
\url{http://www.astro.umontreal.ca/~bergeron/CoolingModels}}. 
WD\,J121856.18$+$254557.18 (WD\,1218) conforms to a white dwarf mass of 0.7--0.9~M$_\sun$ cooling 
for 500~Myr, being consistent with the theory either of \citet{sal10} or of \citet{tre11}, so it is likely
a member. The other white dwarf, WD\,J165132.59$+$681720.10 (WD\,1651) has a similar mass, but has
been cooling for much longer, $\ga6.2$~Gyr, hence it should be a field object.  

\begin{figure*}[tb!]
\centering 
\includegraphics[angle=0, width=1.\textwidth]{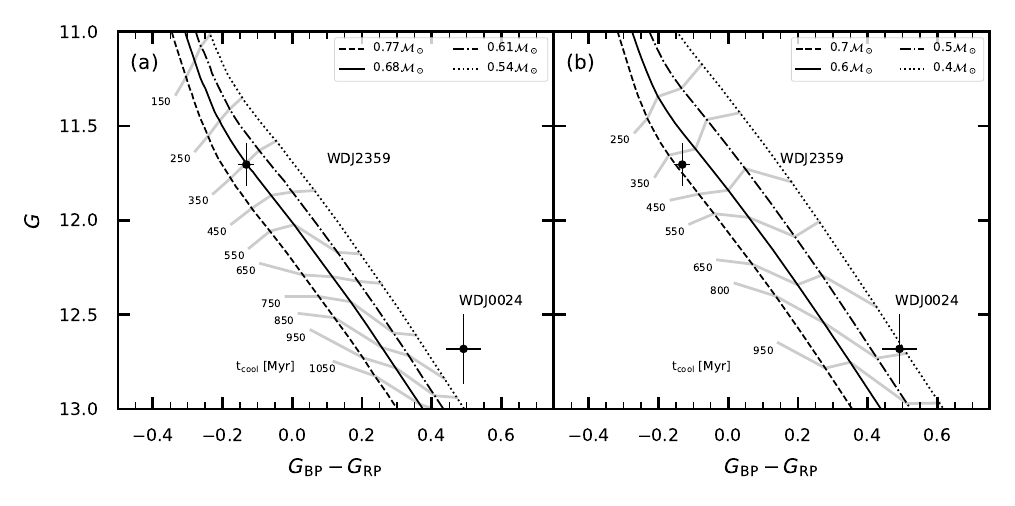}
\caption{The same as Figure~\ref{fig:wd_coma} but for Blanco\,1.  
        }
\label{fig:wd}
\end{figure*}

In the line of sight to Blanco\,1, two white dwarfs, WD\,J235956.52 $-$222103.82 (WD\,2359)  
and WD\,J002421.48 $-$262947.38 (WD\,0024) are identified as possible members as per distance and 
motion; see Figure~\ref{fig:gaiavsstargo}. While WD\,2359 is consistent with being $\sim0.7$~M$_\sun$ 
with a cooling timescale of $\sim350$~Myr, WD\,0024 fits to a mass of 0.4~M$_\sun$ cooling for
$\ga1$~Gyr; see Figure~\ref{fig:wd}. Both two white dwarfs, therefore, are too old to be members  
of Blanco\,1, so cannot be used for age reference of the cluster.  

%%%%%%%%%%%%%%%%%%

%%%%%%%%%%%%%%%%

\begin{figure*}[tb!]
\centering 
\includegraphics[angle=0, width=\textwidth]{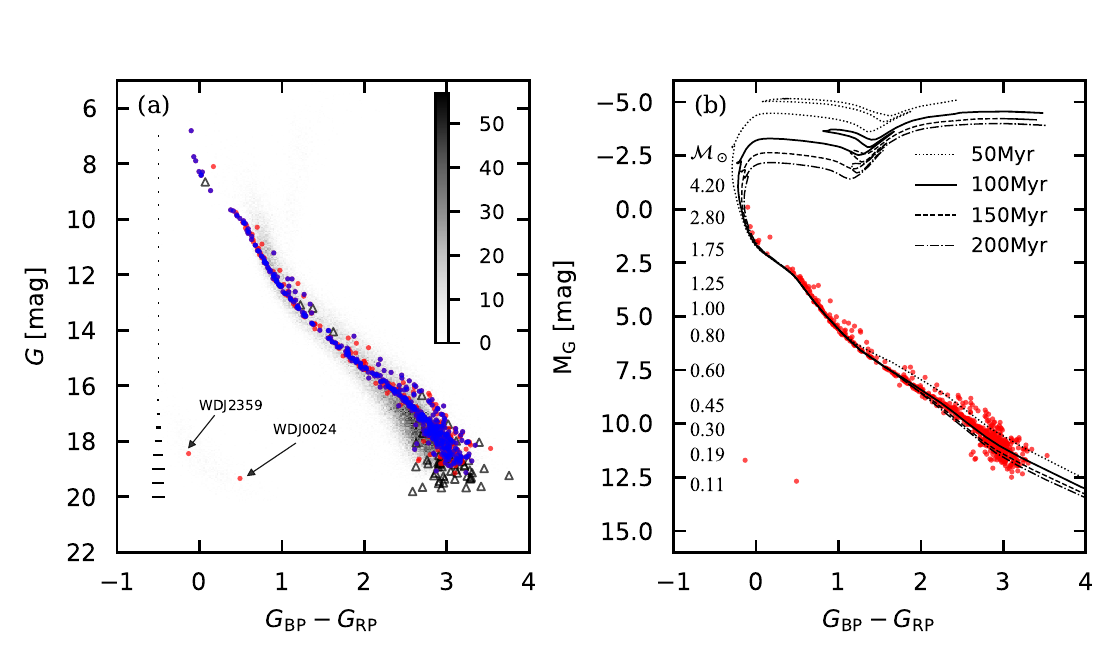}
\caption{The color-magnitude diagrams of {\it Gaia}/DR2 $G$ versus $G_{BP} - G_{RP}$ 
        for Blanco\,1.   
        (a) The observed {\it Gaia} $G$ magnitudes are plotted, with all the stars in Sample~I 
            shown as a density  map in gray, whereas the member candidates are represented by 
            red dots. Typical photometric errors in the color $G_{BP} - G_{RP}$ are represented 
            as horizontal lines on the left. The symbols are the same as Figure~\ref{fig:gaiavsstargo}; 
            that is, blue dots are candidates also found by \citet{gai18b}. The member candidates 
            found only by \citet{gai18b} but not by us are shown as black open triangles.
        (b) The absolute M$_{G}$ magnitudes are plotted after the distance of each member candidate 
            is taken into account. The red dots mark our member candidates. PARSEC isochrones of 
            50, 100, 150, and 200~Myr with solar metallicity and zero extinction are over-plotted. 
            Stellar masses per the 100~Myr isochrone are indicated. 
        }
\label{fig:cmd}
\end{figure*}

Figure~\ref{fig:cmd} exhibits the {\it Gaia}/DR2 color-magnitude diagram (CMD) of Blanco\,1 
in the observed apparent $G$-band magnitudes, and then in absolute M$_{G}$ magnitudes, versus 
the $G_{BP}-G_{RP}$ color after adjusting the distance of each member candidate. A set of 
PARSEC v1.2S isochrones \citep{wei18,che14,tan14,che15}, adopting solar metallicity \citep{for05} 
and no extinction, are also plotted. A 100-Myr isochrone gives an overall satisfactory fit to the 
upper main sequence plus the lower part of the CMD, the latter being low-mass stars still in the 
pre-main sequence phase.  Our sample contains no post-main sequence members, so provides no accurate
age estimate. In this work, we hence adopt an age of 100~Myr for subsequent discussion. 

%---------------------------------------------------------------------------------------%
%---------------------------------------------------------------------------------------%
%sec:morphology
\subsection{The Cluster Shape --- Line of Sight Elongation}\label{sec:z_elong}

\begin{figure*}[tb!]
\centering 
\includegraphics[angle=0, width=\textwidth]{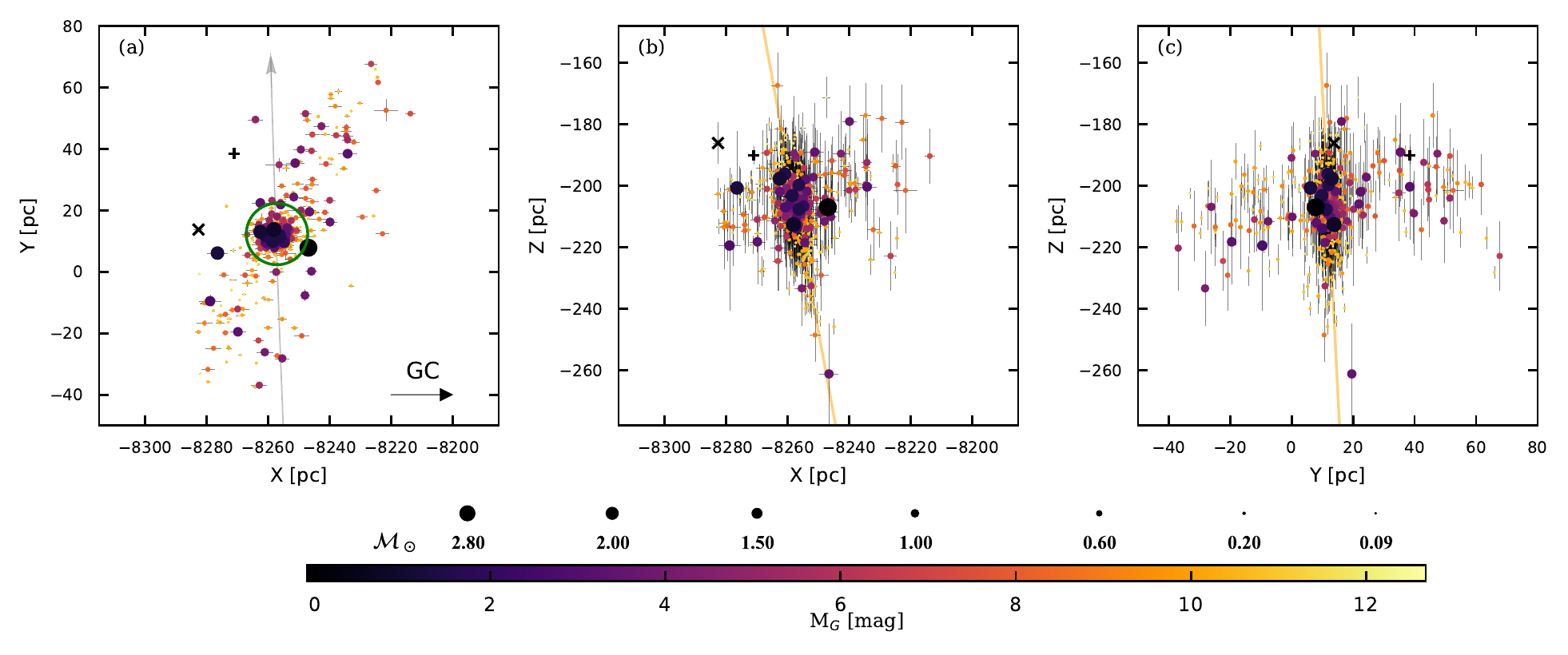}
\caption{
        Our 644 member candidates in Galactocentric Cartesian 
        (a) $X$-$Y$ coordinates, 
        (b) $X$-$Z$ coordinates, and 
        (c) $Y$-$Z$ coordinates. 
        The color shade of the solid circles represents the M$_{G}$~mag and and the size of 
        the symbol depicts the estimated stellar mass.
        The green circle in (a) indicates the 10~pc tidal radius of the cluster,  the grey arrow in (a)
        marks the orbital motion, and the orange lines in (b) and (c) indicate the lines of sight. The
        two white dwarfs are separately marked (a plus sign for WDJ2359, and a cross sign for WDJ0024).
        }
\label{fig:dis_xyz}
\end{figure*}

Blanco\,1 displays an elongation, as evidenced in Figure~\ref{fig:dis_xyz}~(a), in the $X$-$Y$ plane 
as well as a stretch along the Z-axis.  As discussed below, the extension in the $X$-$Y$ plane, which 
happens to be the sky plane, is real, whereas that along the $Z$ direction, i.e., in our line of sight, is likely not.  

\citet{gai18b} reported that with the {\it Gaia}/DR2 data, after certain quality cuts, similar 
to what described in Section~\ref{sec:gaia}, the members of star clusters within a heliocentric
distance of 250~pc, using the Hertzsprung-Russel (HR) diagram analysis, would have $\varpi$ and 
PMs \emph{ ``sufficiently accurate ... to do a 3D reconstruction of each cluster''}. Even though
the errors in parallax measurements $\Delta\varpi$ have a symmetric distribution function, the 
reciprocal function $1/\varpi$, the expected value of which is used in distance computation, has 
an asymmetric distribution, leading to a bias in the distance estimate. We present in 
Appendix~\ref{sec:mc} a Monte Carlo analysis of how the errors $\Delta \varpi$ contribute to the 
evaluation of the $X$,$Y$, and $Z$ coordinates. This artificial elongation, always along the line 
of sight, happens to be nearly in the $Z$ axis for Blanco\,1 (c.f., Figure~\ref{fig:dis_xyz}~(b) 
and (c)).  Note that even for a nearby cluster like Blanco\,1, $\varpi \sim 4$~mas (or $\sim 250$~pc),
a typical $\Delta \varpi$ of 0.24~mas corresponds to a noticeable stretching as large as $\sim15$~pc 
(see Table~\ref{tab:error}).

In the study by \citet{gai18b} on the star cluster HR diagrams, such a bias in distance determination 
is evidenced, as every cluster is seen stretched along the line of sight. Using the method purposed by 
\citet{mad99}, with $\Delta \varpi$, PMs, and a kinematic model on the basis of the convergent point 
method, an estimate of $\varpi$ with an improved precision is afforded. \citet{gai17} demonstrated how
the parallax values would have the errors improved by a factor of two to three better than the observed 
error, and applied this technique to identify members in the solar neighborhood, including Blanco\,1.

\begin{figure}[tb!]
\centering 
\includegraphics[angle=0, width=\columnwidth]{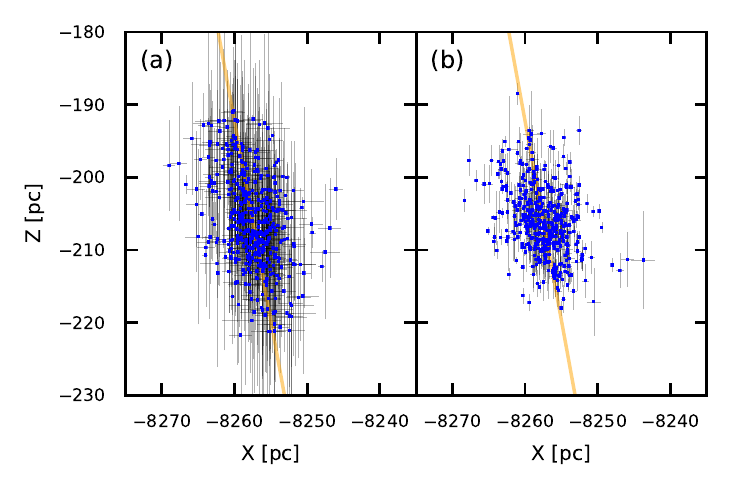}
\caption{ The same as Figure~\ref{fig:dis_xyz} (b) but only
        for the 427 candidates 
        common between this study and that of \citet{gai18b}. 
        (a) The coordinates and their errors are computed from the observed {\it Gaia}/DR2 data.  
        (b) The coordinates and their errors are computed from the ``improved parallax'' given by 
        \citet{gai18b}. In each case, the orange line indicates the line of sight.       }
\label{fig:gdr2}
\end{figure}

Despite all the corrections, the line-of-sight elongation still exists. 
%Of the 489 member candidates 
%of Blanco\,1 found by \citet{gai18b}, 427 are also in our list of candidates. 
Figure~\ref{fig:gdr2} plots the 427 common candidates in Galactocentric Cartesian coordinates derived 
from the {\it Gaia}/DR2 data including the corresponding errors, and derived with the ``improved parallax''
given by \citet{gai18b}. While the situation is partially mitigated, the cluster still appears elongated
along the line of sight.  At the moment, we could not distinguish the level of distortion of the cluster
shape due to this parallax bias, from that due to a possible genuine tidal stretch by the Galactic plane. 

%Of the 62 candidates found by \citet{gai18b} but not included in our member list, %they are mostly proper motion outliers (see Figure~\ref{fig:gaiavsstargo}), either %field stars in the foreground/background with smaller/larger proper motions %(thereby having slower/faster space motion) than member stars.

%---------------------------------------------------------------------------------------%

\subsection{The Cluster Shape --- Tidal Tails }\label{sec:tail}

The tidal radius of a star cluster in the solar neighborhood is computed via \citep{pin98}
\begin{equation} 
  r_t=(\frac{GM_{\rm C}}{2(A-B)^2})^{\frac{1}{3}}, 
\end{equation}
where $G$ is the gravitational constant, $M_{\rm C}$ is the total mass of the cluster, 
and $A$ and $B$ are the Oort constants, 
$A=15.3\pm0.4$~km~s$^{-1}$~kpc$^{-1}$, $B=-11.9\pm0.4$~km~s$^{-1}$~kpc$^{-1} $\citep{bov17}.  
With $M_{\rm C} = 348\pm32$~M$_\sun$ (see Section~\ref{sec:mf}), we estimate the tidal radius 
of Blanco\,1 to be $10.0\pm 0.3$~pc ($\sim2.4\degr$, marked as the green circle in 
Figure~\ref{fig:dis_xyz}~(a)). This radius is close to that of 10.4~pc derived by \citet{mor07}.  
In comparison, \citet{mer08} derived a radius of 7.9~pc with a cluster mass of 160~M$_\sun$,
and \citep{pla11} concluded a $\sim 6$~pc radius with $\sim 300$~M$_\sun$. 
The inconsistency is mainly caused by different cluster mass derived in different studies, or 
by using different formulae to calculate the tidal radius. For example, \citet{pla11} used 
the method by \citet{koz95} that considered the Galaxy mass and the galactocentric distance of
Blanco\,1, rather than the Oort constants we used here, which should be more appropriate and 
accurate in solar neighborhood.

Cluster parameters are derived by member candidates within the cluster's tidal radius, with 
$G$~mag ranging between 9 and 15~mag, and with $\Delta \varpi < 0.5$~mas (see Table~\ref{tab:error}),
resulting in the cluster center of R.A.=$00\fdg7563$, Decl.=$-29\fdg8433$, and $\varpi$=4.2\,mas 
($\sim$ 238.1\,pc), corresponding to Cartesian Galactocentric coordinates of 
$(X,Y,Z) = (-8257.1, +12.3, -207.1)$~pc. The cluster has an average PM of 
$(\mu_\alpha \cos\delta, \mu_\delta) = (+18.7\pm 0.4, +2.6\pm 0.5)$~mas~yr$^{-1}$.
 By adopting the average RV given in Section~\ref{sec:selection}, RV=$6.1\pm1.1$~km~s$^{-1}$,
the average space motion for the cluster relative to the Galactic center 
is $(U,V,W) = (-7.7, +225.6, -2.9)$~km~s$^{-1}$.  The epicyclic motion now brings Blanco\,1 
below and moving away from the plane, and at the same time away from the Galactic center about 
the circular orbit (Local Standard of Rest).  
Subsequent analysis is based on these revised cluster parameters. 
%---------------------------------------------------------------------------------------%

Tidal stripping depends on the mass and also on the age of a cluster.
Member candidates inside the tidal radius (488) outnumber those outside (156) (inside/outside)
by about a factor of 3. This contrasts either with the $\sim700$~Myr old clusters, Coma Berenices,
which has an inside/outside ratio of 0.6 \citep{tan19}, or with the $\sim800$~Myr old, but more massive cluster, Hyades, which has a ratio of 1.1 \citep{ros19a}
\footnote{\citet{ros19a} adopted the cluster boundary as 2 tidal radii. The inside/outside ratio 
would be smaller if the boundary is taken to be one tidal radius, as is the case for 
Coma Berenices and in Blanco\,1.}.  
Blanco\,1 is relatively young and appears to be more dynamically bound.

%---------------------------------------------------------------------------------------%

The orbital motion of Blanco\,1 depicted in Figure~\ref{fig:dis_xyz}~(a) (as a grey arrow) is 
computed based on the average position and median velocity via the Python \texttt{galpy} 
package \citep{bov15} \footnote{The Galactic gravitational potential used here is 
``\texttt{MWPotential2014}'', a model that comprises the bulge, disk, and halo. Parameters of the 
model were fitted to published dynamical data of the Milky Way.}. A leading tail (to the positive 
$Y$-axis direction, green open circles in Figure~\ref{fig:gaiavsstargo}~(a)) and a trailing tail 
(to the negative $Y$-axis direction, red open circles), each with an extension of 50--60~pc projected
in the plane of the sky, are revealed. 

%Can the angle between the orbital motion of the cluster and that of
%the tidal tails (e.g., Fig. 9(a)) tell us something more about the
%orbit of the cluster?

%---------------------------------------------------------------------------------------%

 The tails may be caused by tidal forces from a nearby massive object, by disk crossing, or by differential rotation in the disk. It is unlikely due to the first mechanism because there are 
no obvious tidal sources in the vicinity of Blanco\,1, such as a giant molecular cloud or a star 
cluster. On the other hand, the compressive shocking during a disk crossing would result in a 
somewhat splashdown morphology, unlike what is observed in this cluster.  
Blanco\,1 is directly above the Sun vertical to the plane so shares the same differential 
rotation as the Sun. Given the Oort constant $A$, 
which measures the shear motion in the Galactic disk at the location of the Sun, the Galactic 
differential rotation across the span of $\sim100$~pc of the Blanco\,1 tails would then lead to
a velocity difference comparable to the internal velocity dispersion of $\la 1$~km~s$^{-1}$ typical 
in open clusters, signifying the potential importance of differential rotation in cluster disruption.
The stretching in this case is aligned with the Lagragian boundary between the competing internal 
gravitational potential and the external shearing potential.   

\begin{figure}[tb!]
\centering
\includegraphics[angle=0, width=1.\columnwidth]{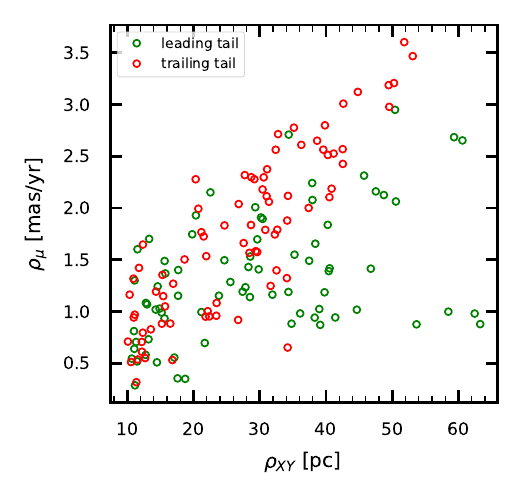}
\caption{
        The distance from the cluster center on the $X$-$Y$ plane versus the 
        difference of the stellar PM with respect to the PM center for the leading (green) or trailing (red) (same as in Figure~\ref{fig:gaiavsstargo}) tail.  In the trailing tail, stars located further from the cluster center stream away faster.  In the leading tail, no such 
        a correlation is evidenced.
                }
\label{fig:Hubble_flow}
\end{figure}

Whatever the mechanism, the distorted cluster would bulge or stretch symmetrically about the cluster,
but the unbound stars should trace a stellar stream primarily in the downstream of the cluster's 
motion. The Blanco\,1 tails stand out clearly not only in space, but also in kinematics. The extended 
``PM tail'' located around $\mu_\delta < 2$~mas~yr$^{-1}$ in Figure~\ref{fig:gaiavsstargo}~(b) 
corresponds to the leading tail in space (Figure~\ref{fig:gaiavsstargo}~(a)), whereas the ``PM tail'' 
located around $\mu_\delta > 4$~mas~yr$^{-1}$ corresponds to the trailing tail. Moreover, in the 
trailing tail, which happens to be in the Blanco\,1 orbital downstream direction, stars further 
distant from the cluster center stream away faster, in support of the scenario of escaping members 
from the cluster. Such a ``Hubble flow'' is not evident in the leading tail; see Figure~\ref{fig:Hubble_flow}.  

Any star cluster in the disk should continue to disintegrate, hence the debris/tail structure 
should be ubiquitous. Incidentally, Blanco\,1, seen toward the Galactic South Pole, and Coma Berenices,
projected near the Galactic North Pole, form an interesting comparison pair that render a face-on view 
of a cluster (tails) in the plane. While each of the two spatial tails of Blanco\,1 corresponds to a 
distinct extended feature in PM, in Coma Berenices, also with two spatial tails, only one PM tail 
is clearly discerned because of the projection effect. 

%fig11

%---------------------------------------------------------------------------------------%
%---------------------------------------------------------------------------------------%

\subsection{The Cluster Mass Function}\label{sec:mf}

\begin{figure*}[tb!]
\centering 
\includegraphics[angle=0, width=0.85\textwidth]{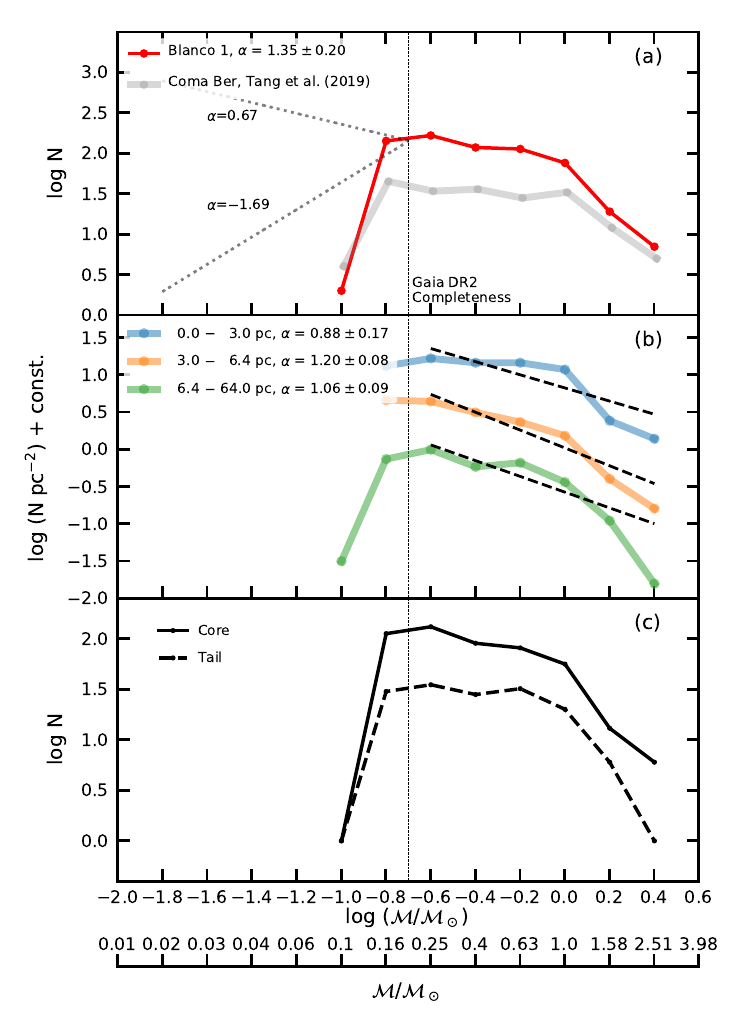}
\caption{
	    (a) The present-day mass function of Blanco\,1, depicted as the red curve, adopting 
	        a 100-Myr PARSEC isochrone. From mass 0.25~M$_\sun$ to 2.51~M$_\sun$, the mass 
	        function slope is $\alpha = 1.35\pm0.20$. The dotted gray lines are for low-mass
	        objects not included in this study, and taken from \citet{mor07} with $\alpha = 0.67$, 
	        and from \citep{tan18} with $\alpha = -1.69$.  
	    (b) The surface number density mass functions on the $X$-$Y$ plane. In each annulus, 
	        $\alpha$ is computed from 0.25~M$_\sun$ to 2.51~M$_\sun$, with a bin size of 
	        $\log$(M/M$_\sun$)=0.2. 
	    (c)  The present-day mass function for members inside (solid line), and outside (dashed line)
	        the tidal radius.
	    % The $X$-value of each dot represents the central value in each bin. 
        }
\label{fig:mf}
\end{figure*}

We estimated the stellar mass using the mass-magnitude relation per the 100~Myr PARSEC isochrone.
 Our membership selection, and hence the subsequent mass function of the cluster, are limited by 
the {\it Gaia}/DR2 brightness completeness of $G~\sim~18.5$, corresponding to $\sim0.2$~M$_\sun$.
The mass function (MF) of the cluster is shown as Figure~\ref{fig:mf}~(a). In the 
form of $dN/dm \propto m^{-\alpha}$, where $N$ is the number of members within the mass bin $dm$, 
a slope is derived, by linear least-squares fitting from mass 0.25~M$_\sun$ to 2.51~M$_\sun$, 
to be $\alpha = 1.35\pm0.20$.  

For lower masses into the brown dwarf regime, the slope has been 
found to be flatter, with $\alpha = 0.69\pm0.15$ by \citet{mor07} in the mass range between 
0.03~M$_\sun$ to 0.6~M$_\sun$. Later, \citet{cas12}, after excluding non-members from the 
\citet{mor07} sample plus their own observations, determined $\alpha = 0.93\pm0.11$ in the same
mass range.  

With a similar age to Blanco\,1, Pleiades \citep[$\sim$ 125\,Myr,][]{sta98} has often been 
considered as a scaled up ``twin'' cluster of Blanco\,1. \citet{cas07} derived a slope of 
$\alpha = 0.35\pm0.31$ for Pleiades in the mass range 0.02--0.06~M$_\sun$. \citet{sta07} inferred 
$\alpha \sim 1$ for low-mass stars ranging from 0.2~M$_\sun$ to 0.5~M$_\sun$. Despite a possibly 
high contamination rate and incompleteness, either of these studies shows an increase into the 
substellar population. On the other hand, the relatively older 700~Myr cluster Coma Berenices has an MF
turned around $\sim0.3$~M$_\sun$, and has a slope of $\alpha = -1.69\pm0.14$ in the low-mass 
end 0.06--0.3~M$_\sun$ \citep{tan18}. 
 This can be seen as evidence of the evolution of the MF. Moreover, in 
Figure~\ref{fig:mf}~(a), comparing the MF of Blanco1 and Coma Ber \citep[grey line,][]{tan19}. 
It is clear that in the high mass range (1.0~M$_\sun$ to 2.5~M$_\sun$), both clusters have similar
fractions of members, but the older Coma Berenices is more depleted in the lower mass end.

The total {\em detected} mass of Blanco\,1 was estimated by adding up the masses of member 
candidates having mass $> 0.2$~M$_\sun$, amounting to 285~M$_\sun$.  To estimate the member 
population below 0.2~M$_\sun$, we exercised two limiting cases.  One is by the flatter 
slope of $\alpha \sim 0.67$ from \citet{mor07}, likely an over estimate, to integrate from 
0.02~M$_\sun$ to 0.16~M$_\sun$, leading to a total cluster mass of 380~M$_\sun$. The lower limit 
is for $\alpha = -1.69$ for Coma Berenices \citep{tan18}, giving a total mass of 316~M$_\sun$.  
An average of $348\pm32$~M$_\sun$ is adopted for the following dynamic analysis.

%---------------------------------------------------------------------------------------%
%---------------------------------------------------------------------------------------%

\subsection{Mass Segregation}

A cluster with a high level of mass segregation often is readily discernible in the 
density profile \citep{hil97,wan14}, or the MF inside different annuli \citep{pan13,tan19}. 
These methods, nonetheless, are effective only when a large number of stars are segregated. 
\citet{gou04} reported different levels of mass segregation among the young clusters in the
Large Magellanic Cloud and the Small Magellanic Cloud via the two methods mentioned above with 
different number of mass bins, and with different bin sizes. Figure~\ref{fig:mf}~(b) displays 
the MF of Blanco\,1 in three annuli, and there is no obvious increase in the slope from inside-out,
suggesting no mass segregation.
 This conclusion can also be made from Figure~\ref{fig:mf}~(c), that the shape of the MF 
of the core members (solid line) and the tail members (dashed line) are indistinguishable.

Alternatively, we adopted the $\Lambda$ method based on the minimum spanning tree (MST) algorithm
developed by \citet{all09a}. An application of the $\Lambda$ method to quantify the level of mass 
segregation in a young star cluster can be found in \citet{pan13}. In brief, the $\Lambda$ method 
compares the average distance among the $N$ most massive members ($l_{\rm massive}$) of the cluster,
to that of the $N$ random members ($l_{\rm normal}$). If $l_{\rm massive}$ is smaller, the cluster
is mass segregated to its $N$th massive stars. Note that the distance between each pair of stars is 
calculated by the MST method, and $l_{\rm normal}$ is the average length of 100 random sets. 
The significance of the mass segregation, parameterized as the ``the mass segregation ratio'' 
($\Lambda_{\rm MSR}$) is defined \citep{all09a} as 
\begin{equation}
 \Lambda_{\rm MSR} = \frac{\langle l_{\rm normal} \rangle}{l_{\rm massive}} 
   \pm \frac{\sigma_{\rm normal}}{l_{\rm massive}}, 
\end{equation}
with $\sigma_{\rm normal}$ being the standard deviation of the 100 different sets of $l_{\rm normal}$.

\begin{figure}[tb!]
\centering
\includegraphics[angle=0, width=1.\columnwidth]{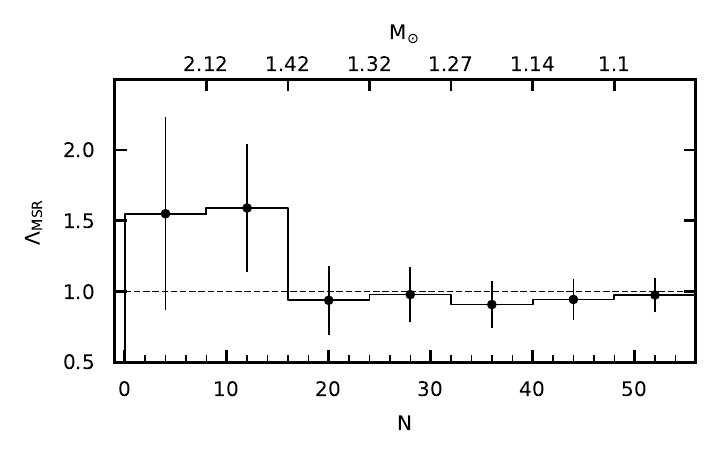}
\caption{
        The ``the mass segregation ratio'' ($\Lambda_{\rm MST}$) for the 56 most 
        massive member candidates with a bin size of 8 stars.
        The dashed line of $\Lambda_{\rm MST} = 1$ indicates no mass segregation, and the
        larger the $\Lambda_{\rm MST}$ is, the more significant the mass segregation is. 
        Error bars are from 100 different realizations of $\emph{l}_{\rm normal}$.  
         The fewer number of more massive stars contributes to always larger 
        stochastic errors $\langle l_{\rm normal} \rangle$ for smaller N$_{\rm MST}$.
        %The corresponding 
        %stellar masses of $N= 8, 16, 24, 32, 45, 48$ are also labeled.
        }
\label{fig:mst}
\end{figure}

 Figure~\ref{fig:mst} presents $\Lambda_{\rm MSR}$ for Blanco\,1, which suggests possible 
mass segregation ($\Lambda_{\rm MSR} > 1.5$) for members $\gtrsim 1.4$~M$_\sun$, i.e., among the most
massive members in our sample, but not for less massive members. Our results cannot be reconciled with 
what \citet{mor07} claimed that mass segregation occurs in this cluster for 0.09--0.6~M$_\sun$, and also 
for the substellar population 0.03--0.08~M$_\sun$. 

As a consequence of energy equipartition in two-body encounters, mass segregation occurs  
faster among (fewer) more massive stars, and proceeds toward lower masses as a star cluster ages.  
The time for a cluster to segregate ($t_{\rm seg}$) down to mass $M$ is 
\citep{spi87, all09b, pan13},
\begin{equation}
  t_{\rm seg}(M) \sim \frac{\langle m \rangle}{M} t_{\rm relax} 
      = \frac{\langle m \rangle}{M} \frac{N}{8\,\ln N} t_{\rm cross}, 
\end{equation}
where $\langle m \rangle$ is the average stellar mass in the cluster, 
$N$ is the number of members of the zero-age cluster, and $t_{\rm cross}$ is the 
crossing time, which is the size of the cluster ($D$) divided by the cluster's velocity 
dispersion ($\sigma$). 

 In analogy between Pleiades and Blanco\,1, and assuming a 10\% (a lower limit) loss
of members at an age of 100~Myr in Blanco\,1 as in Pleiades \citep{mor07,ada02,mor04},  
the initial members in Blanco\,1 would be $N\approx 715$. Taking $D$ to be the tidal radius 
10~pc, $\sigma \sim 1$~km~s$^{-1}$ \citep{gon09}, and $\langle m \rangle \sim 0.5$~M$_\sun$,
Equation~(3) suggests $t_{\rm seg}\approx 50$~Myr for mass 1.4~M$_\sun$. Therefore, Blanco\,1  
is at just the right age to allow for mass segregation for members down to 1.4~M$_\sun$.

%{\bf Considering more amount members loss at an age of 100~Myr in Blanco\,1, 
%the according $t_{\rm seg}$ becomes larger for mass 1.4~M$_\sun$. 
%---------------------------------------------------------------------------------------%
%---------------------------------------------------------------------------------------%

\section{Summary} \label{sec:summary}

With {\it Gaia}/DR2 photometry and astrometry, we used 5 parameters (Galactic position $X$,
$Y$, and $Z$, and proper motions $\mu_\alpha \cos\delta, \mu_\delta$) to secure a list of 
644 member candidates of the star cluster Blanco\,1. The tidal structure on the $X$-$Y$ plane,
with a leading tail extending $\sim60$~pc and a trailing tail of $\sim50$~pc, each 5--6 times 
the size of the cluster's tidal radius of $10.0\pm 0.3$~pc, has been detected for the first 
time for this cluster. The extended shape along the line of sight, roughly in the $Z$ direction 
for Blanco\,1, is found to be artificial, caused by a systematic parallax bias.  

The member candidates have a central position of R.A.=$00\fdg7563$, Decl.=$-29\fdg8433$, an
average parallax of $\varpi=4.2$~mas (i.e., $\sim238.1$~pc), proper motion of 
$(\mu_\alpha \cos\delta, \mu_\delta) = (+18.7\pm0.4, +2.6\pm0.5)$~mas~yr$^{-1}$, and radial 
velocity of RV=$6.1\pm1.1$~km~s$^{-1}$. The corresponding Cartesian Galactocentric coordinates 
are $(X,Y,Z) = (-8257.1, +12.3, -207.1)$~pc, 
with the average space motion as $(U,V,W)$=$(-7.7, +225.6, -2.9)$ km~s$^{-1}$ relative 
to the Galactic center.
The revised list of member candidates is consistent with an age of 100~Myr, and gives a total
cluster mass of $348\pm32$~M$_\sun$. The cluster has a mass function of a slope of
$\alpha = 1.35\pm0.2$ for mass from 0.25~M$_\sun$ to 2.51~M$_\sun$, with mass segregation 
evidenced only for members more massive than 1.4~M$_\sun$,  suggesting an initial dynamical 
disintegration.

%$(U,V,W) = (-18.8, -6.6, -10.1)$~km~s$^{-1}$
%---------------------------------------------------------------------------------------%
%---------------------------------------------------------------------------------------%

\acknowledgments
We thank the anonymous referee for constructive comments that significantly improve the 
quality of the paper.  
Y.Z and J.Z.L acknowledge the financial support of National Natural Science Foundation 
of China (No.~11661161016). Y.Z is also supported by the Youth Innovation Promotion 
Association CAS (No.~2018080), 2017 Heaven Lake Hundred-Talent Program of Xinjiang Uygur 
Autonomous Region of China, and the program of Tianshan Youth (No.~2017Q091).
S.Y.T, and W.P.C. acknowledge the financial support of the grants MOST 106-2112-M-008-005-MY3 
and MOST 105-2119-M-008-028-MY3. X.Y.P. expresses gratitude for support from the Research
Development Fund of Xi’an Jiaotong Liverpool University (RDF-18-02-32) and the financial 
support of two grants of National Natural Science Foundation of China, Nos.~11673032 and 11503015. 
We thank Prof. Dr. M.~B.~N. Kouwenhoven for making available the MST code. This work has made use 
of data from the European Space Agency (ESA) mission {\it Gaia} (\url{https://www.cosmos.esa.int/gaia}), 
processed by the {\it Gaia} Data Processing and Analysis Consortium (DPAC,
\url{https://www.cosmos.esa.int/web/gaia/dpac/consortium}). 
 
%---------------------------------------------------------------------------------------%
%---------------------------------------------------------------------------------------%
\software{  Astropy \citep{ast13,ast18}, 
            SciPy \citep{mil11},
            galpy \citep{bov15}, and \textsc{StarGO} \citep{yua18}
}
%---------------------------------------------------------------------------------------%
%---------------------------------------------------------------------------------------%

%-------------------------------------------------------------------------------------------%
%*******************************************************************************************%
\appendix

\section{Gaia\,DR2 Quality Cuts}\label{sec:qulitycut}

The quality cuts on the {\it Gaia}\,DR2 data performs in this study are as follow: \\
\texttt{parallax\_over\_error}$>$10,                \\
\texttt{phot\_g\_mean\_flux\_over\_error}$>$10,     \\
\texttt{phot\_rp\_mean\_flux\_over\_error}$>$10,    \\
\texttt{phot\_bp\_mean\_flux\_over\_error}$>$10,    \\
\texttt{visibility\_periods\_used}$>$5,             \\
\texttt{astrometric\_excess\_noise}$<$1,            \\
\texttt{phot\_bp\_rp\_excess\_factor}$<$1.3+0.060$ \times
(\texttt{phot\_bp\_mean\_mag}$-$\texttt{phot\_rp\_mean\_mag})^2$, and   \\
\texttt{phot\_bp\_rp\_excess\_factor}$>$1.0+0.015$ \times
(\texttt{phot\_bp\_mean\_mag}$-$\texttt{phot\_rp\_mean\_mag})^2$.

%*******************************************************************************************%
\section{Distance Error Estimation With Monte Carlo Method}\label{sec:mc}

The $\Delta \varpi$ of each object in {\it Gaia}\,DR2 is related to the brightness of each 
object, e.g., the fainter a star is the larger the $\Delta \varpi$ will be. Hence, we take the
mean $\Delta \varpi$ in different $G$ magnitude ranges as the typical parallax errors 
(column~2 in Table.~\ref{tab:error}). 

Same $\Delta \varpi$ will have different levels of impact for stars in different distances. 
Thus, we demonstrate here with $\varpi =$ 10, 4, and 2~mas. By taking the $\varpi$ as mean and 
$\Delta \varpi$ as one $\sigma$, we re-generate 100,000 mock $\varpi$ values. These values are 
then be taken, 1/$\varpi$, to get 100,000 distances and with their stander deviation to be the
estimated distance error (see Table.~\ref{tab:error}).

\begin{deluxetable}{cC C C C}
\tablecaption{Distance Errors for Different $G$ Magnitudes at 
                Different Distances \label{tab:error}
             }
\tabletypesize{\scriptsize}
\tablehead{
    \colhead{ } &   \colhead{ } &
    \colhead{10\,mas (100\,pc)} & \colhead{4\,mas (250\,pc)} & \colhead{2\,mas (500\,pc)} \\
    \colhead{ } &   \colhead{ } &
    \colhead{Coma Berenices}          &\colhead{Blanco\,1}         & \colhead{NGC\,2422}        \\
    \cline{3-5}
	\colhead{$G$}               & \colhead{$\Delta \varpi$}  & \colhead{$\Delta$ Dist.}   &
	\colhead{$\Delta$ Dist.}    & \colhead{$\Delta$ Dist.}                                \\
	\colhead{(mag)}             & \colhead{(mas)}            & \colhead{(pc)}             &
	\colhead{(pc)}              & \colhead{(pc)}                                          \\
	\colhead{(1)}             & \colhead{(2)}            & \colhead{(3)}             &
	\colhead{(4)}              & \colhead{(5)}          
	 }
%\colnumbers
\startdata
 5 -  6 &  0.07 &   0.7 &    4.7 &  18.8 \\ 
 6 -  7 &  0.10 &   1.0 &    6.4 &  26.0 \\ 
 7 -  8 &  0.06 &   0.6 &    3.8 &  15.2 \\ 
 8 -  9 &  0.06 &   0.6 &    3.9 &  15.8 \\ 
 9 - 10 &  0.05 &   0.5 &    3.3 &  13.1 \\ 
10 - 11 &  0.05 &   0.5 &    3.1 &  12.5 \\ 
11 - 12 &  0.04 &   0.4 &    2.8 &  11.2 \\ 
12 - 13 &  0.04 &   0.4 &    2.8 &  11.1 \\ 
13 - 14 &  0.03 &   0.3 &    1.9 &   7.5 \\ 
14 - 15 &  0.04 &   0.4 &    2.3 &   9.2 \\ 
15 - 16 &  0.06 &   0.6 &    3.5 &  14.1 \\ 
16 - 17 &  0.09 &   0.9 &    5.6 &  22.5 \\ 
17 - 18 &  0.14 &   1.4 &    9.1 &  36.8 \\ 
18 - 19 &  0.24 &   2.4 &   14.9 &  62.4 \\ 
\enddata
\tablecomments{Mean distance for Coma Berenices star cluster is 85.5\,pc \citep{tan19}, 
                for Blanco\,1 is 238.1\,pc (this work), 
                and for NGC2422 is 483.3\,pc \citep{gai18}.
            }
\end{deluxetable} 
%-------------------------------------------------------------------------------------------%


\begin{thebibliography}{}

\bibitem[Adams et al.(2002)]{ada02} 
        Adams, T., Davies, M.~B., Jameson, R.~F., et al.\ 2002, \mnras, 333, 547

\bibitem[Allison et al.(2009a)]{all09a} 
        Allison, R.~J., Goodwin, S.~P., Parker, R.~J., et al.\ 2009a, \mnras, 395, 1449

\bibitem[Allison et al.(2009b)]{all09b} 
        Allison, R.~J., Goodwin, S.~P., Parker, R.~J., et al.\ 2009b, \apj, 700, L99

\bibitem[Astropy Collaboration et al.(2013)]{ast13} 
        Astropy Collaboration, Robitaille, T.~P., Tollerud, E.~J., et al.\ 2013, \aap, 558, A33 

\bibitem[Astropy Collaboration et al.(2018)]{ast18}
        Astropy Collaboration, Price-Whelan, A.~M., Sip{\H o}cz, B.~M., et al.\ 2018, \aj, 156, 123

\bibitem[Bailer-Jones et al.(2018)]{bai18} 
        Bailer-Jones, C.~A.~L., Rybizki, J., Fouesneau, M., Mantelet, G., \& Andrae, R.\ 2018, \aj, 156, 58 

\bibitem[Bergeron et al.(2011)]{ber11} 
        Bergeron, P., Wesemael, F., Dufour, P., et al.\ 2011, \apj, 737, 28 

\bibitem[Blanco(1949)]{bla49} 
        Blanco, V.~M.\ 1949, \pasp, 61, 183

\bibitem[Bovy(2015)]{bov15} 
        Bovy, J.\ 2015, \apjs, 216, 29

\bibitem[Bovy(2017)]{bov17} 
        Bovy, J.\ 2017, \mnras, 468, L63 
        
\bibitem[Cargile et al.(2009)]{car09} 
        Cargile, P.~A., James, D.~J., Platais, I.\ 2009, \aj, 137, 3230 

\bibitem[Cargile et al.(2014)]{car14} 
        Cargile, P.~A., James, D.~J., Pepper, J., et al.\ 2014, \apj, 782, 29 

\bibitem[Casewell et al.(2007)]{cas07} %pleiades BD MF slope
        Casewell, S.~L., Dobbie, P.~D., Hodgkin, S.~T., et al.\ 2007, \mnras, 378, 1131

\bibitem[Casewell et al.(2012)]{cas12} 
        Casewell, S.~L., Baker, D.~E.~A., Jameson, R.~F., et al.\ 2012, \mnras, 425, 3112 

\bibitem[Chen, \& Chen(2010)]{che10} 
        Chen, C.~W., \& Chen, W.~P.\ 2010, \apj, 721, 1790

\bibitem[Chen et al.(2004)]{che04} 
        Chen, W.~P., Chen, C.~W., \& Shu, C.~G.\ 2004, \aj, 128, 2306

\bibitem[Chen et al.(2014)]{che14} 
	    Chen, Y., Girardi, L., Bressan, A., et al.\ 2014, \mnras, 444, 2525 

\bibitem[Chen et al.(2015)]{che15} 
	    Chen, Y., Bressan, A., Girardi, L., et al.\ 2015, \mnras, 452, 1068 

\bibitem[de Epstein \& Epstein(1985)]{ded85} 
        de Epstein, A.~E.~A., \& Epstein, I.\ 1985, \aj, 90, 1211 

\bibitem[Dehnen et al.(2004)]{deh04} 
        Dehnen, W., Odenkirchen, M., Grebel, E.~K., et al.\ 2004, \aj, 127, 2753

\bibitem[Ford et al.(2005)]{for05} 
        Ford, A., Jeffries, R.~D., \& Smalley, B.\ 2005, \mnras, 364, 272 

\bibitem[F{\"u}rnkranz et al.(2019)]{fur19} 
        F{\"u}rnkranz, V., Meingast, S., \& Alves, J.\ 2019, \aap, 624, L11 

\bibitem[Gaia Collaboration et al.(2017)]{gai17} 
        Gaia Collaboration, van Leeuwen, F., Vallenari, A., et al.\ 2017, \aap, 601, A19 

\bibitem[Gaia Collaboration et al.(2018a)]{gai18} 
        Gaia Collaboration, Brown, A.~G.~A., Vallenari, A., et al.\ 2018a, \aap, 616, A1 

\bibitem[Gaia Collaboration et al.(2018b) b]{gai18b} 
        Gaia Collaboration, Babusiaux, C., van Leeuwen, F., et al.\ 2018b, \aap, 616, A10

\bibitem[Gonz{\'a}lez, \& Levato(2009)]{gon09} 
        Gonz{\'a}lez, J.~F., \& Levato, H.\ 2009, \aap, 507, 541.

\bibitem[Gouliermis et al.(2004)]{gou04} 
        Gouliermis, D., Keller, S.~C., Kontizas, M., et al.\ 2004, \aap, 416, 137

\bibitem[Heggie, \& Hut(2003)]{heg03} 
        Heggie, D., \& Hut, P.\ 2003, The Gravitational Million-Body Problem: A Multidisciplinary Approach to Star Cluster Dynamics, Cambridge U Press

\bibitem[Hillenbrand(1997)]{hil97} 
        Hillenbrand, L.~A.\ 1997, \aj, 113, 1733

\bibitem[Holberg \& Bergeron(2006)]{hol06} 
        Holberg, J.~B., \& Bergeron, P.\ 2006, \aj, 132, 1221 

\bibitem[Juarez et al.(2014)]{jua14} 
        Juarez, A.~J., Cargile, P.~A., James, D.~J., \& Stassun, K.~G.\ 2014, \apj, 795, 143 

\bibitem[Kozhurina-Platais et al.(1995)]{koz95} 
        Kozhurina-Platais, V., Girard, T.~M., Platais, I., et al.\ 1995, \aj, 109, 672 

\bibitem[Kuzma et al.(2015)]{kuz15} 
        Kuzma, P.~B., Da Costa, G.~S., Keller, S.~C., et al.\ 2015, \mnras, 446, 3297

\bibitem[Lada \& Lada(2003)]{lad03} 
	    Lada, C.~J., \& Lada, E.~A.\ 2003, \araa, 41, 57 

\bibitem[Lindegren et al.(2018)]{lin18} 
        Lindegren, L., Hern{\'a}ndez, J., Bombrun, A., et al.\ 2018, \aap, 616, A2 

\bibitem[Madsen(1999)]{mad99} 
        Madsen, S.\ 1999, ASPC 167, Harmonizing Cosmic Distance Scales in a Post-hipparcos Era,
        eds, Egret, D., and Heck, A., 78

\bibitem[Meingast \& Alves(2019)]{mei18} 
        Meingast, S., \& Alves, J.\ 2019, \aap, 621, L3 

\bibitem[Mermilliod et al.(2008)]{mer08} 
        Mermilliod, J.-C., Platais, I., James, D.~J., Grenon, M., \& Cargile, P.~A.\ 2008, \aap, 485, 95

\bibitem[Micela et al.(1999)]{mic99} 
        Micela, G., Sciortino, S., Favata, F., et al.\ 1999, \aap, 344, 83

\bibitem[Millman et al.(2011)]{mil11} 
        Millman, K. J., Aivazis, M..\ 2011, Computing in Science \& Engineering, 13, 2, 9

\bibitem[Moraux et al.(2004)]{mor04} 
        Moraux, E., Kroupa, P., \& Bouvier, J.\ 2004, \aap, 426, 75

\bibitem[Moraux et al.(2007)]{mor07} 
        Moraux, E., Bouvier, J., Stauffer, J.~R., Barrado y Navascu{\'e}s, D., \& Cuillandre, J.-C.\ 2007, \aap, 471, 499 

\bibitem[Odenkirchen et al.(2001)]{ode01} 
        Odenkirchen, M., Grebel, E.~K., Rockosi, C.~M., et al.\ 2001, \apjl, 548, L165

\bibitem[Odenkirchen et al.(2003)]{ode03} 
        Odenkirchen, M., Grebel, E.~K., Dehnen, W., et al.\ 2003, \aj, 126, 2385
    
\bibitem[Panagi \& O'dell(1997)]{pan97} 
        Panagi, P.~M., \& O'dell, M.~A.\ 1997, \aaps, 121, 213 

\bibitem[Pang et al.(2013)]{pan13} 
        Pang, X., Grebel, E.~K., Allison, R.~J., et al.\ 2013, \apj, 764, 73

\bibitem[Perry et al.(1978)]{per78} 
        Perry, C.~L., Walter, D.~K., \& Crawford, D.~L.\ 1978, \pasp, 90, 81

\bibitem[Platais et al.(2011)]{pla11} 
        Platais, I., Girard, T.~M., Vieira, K., et al.\ 2011, \mnras, 413, 1024 

\bibitem[Pillitteri et al.(2003)]{pil03} 
        Pillitteri, I., Micela, G., Sciortino, S., et al.\ 2003, \aap, 399, 919
        
\bibitem[Pillitteri et al.(2004)]{pil04} 
        Pillitteri, I., Micela, G., Sciortino, S., et al.\ 2004, \aap, 421, 175
        
\bibitem[Pillitteri et al.(2005)]{pil05} 
        Pillitteri, I., Micela, G., Reale, F., et al.\ 2004, \aap, 421, 17

\bibitem[Pinfield et al.(1998)]{pin98} 
        Pinfield, D.~J., Jameson, R.~F., \& Hodgkin, S.~T.\ 1998, \mnras, 299, 955

\bibitem[R{\"o}ser et al.(2019)]{ros19a} 
        R{\"o}ser, S., Schilbach, E., \& Goldman, B.\ 2019, \aap, 621, L2 

\bibitem[R{\"o}ser, \& Schilbach(2019)]{ros19b} 
        R{\"o}ser, S., \& Schilbach, E.\ 2019, \aap, 627, A4

\bibitem[Rybizki et al.(2018)]{ryb18} 
        Rybizki, J., Demleitner, M., Fouesneau, M., et al.\ 2018, \pasp, 130, 74101

\bibitem[Salaris et al.(2010)]{sal10} 
        Salaris, M., Cassisi, S., Pietrinferni, A., et al.\ 2010, \apj, 716, 1241

\bibitem[Spitzer(1987)]{spi87} 
        Spitzer, L. 1987, Dynamical Evolution of Globular Clusters (Princeton, NJ: Princeton Univ. Press)

\bibitem[Stauffer et al.(1998)]{sta98} 
        Stauffer, J.~R., Schultz, G., \& Kirkpatrick, J.~D.\ 1998, \apjl, 499, L199 

\bibitem[Stauffer et al.(2007)]{sta07} %pleiades BD MF slope
        Stauffer, J.~R., Hartmann, L.~W., Fazio, G.~G., et al.\ 2007, \apjs, 172, 663

\bibitem[Tang et al.(2014)]{tan14}
	    Tang, J., Bressan, A., Rosenfield, P., et al.\ 2014, \mnras, 445, 4287

\bibitem[Tang et al.(2018)]{tan18} 
        Tang, S.-Y., Chen, W.~P., Chiang, P.~S., et al.\ 2018, \apj, 862, 106

\bibitem[Tang et al.(2019)]{tan19} 
        Tang, S.-Y., Pang, X., Yuan, Z., et al.\ 2019, \apj, 877, 1

\bibitem[Tremblay et al.(2011)]{tre11} 
        Tremblay, P.-E., Bergeron, P., \& Gianninas, A.\ 2011, \apj, 730, 128 

\bibitem[Wang et al.(2014)]{wan14} 
        Wang, P.~F., Chen, W.~P., Lin, C.~C., et al.\ 2014, \apj, 784, 57

\bibitem[Weiler(2018)]{wei18} 
        Weiler, M.\ 2018, \aap, 617, A138

\bibitem[Westerlund et al.(1988)]{wes88} 
        Westerlund, B.~E., Garnier, R., Lundgren, K., et al.\ 1988, \aaps, 76, 101
    
\bibitem[Yuan et al.(2018)]{yua18} 
        Yuan, Z., Chang, J., Banerjee, P., et al.\ 2018, \apj, 863, 26

\end{thebibliography}
\end{document}